\documentclass[11pt]{article}
\usepackage{hyperref}
\pdfoutput=1
\begin{document}
\title{Airflow in a Multiscale Subject-Specific \\ Breathing Human Lung Model}
\author{Jiwoong Choi\textsuperscript{1}, Youbing Yin\textsuperscript{1}, Eric A. Hoffman\textsuperscript{2}, Merryn H. Tawhai\textsuperscript{3}, \\
and Ching-Long Lin\textsuperscript{1} \\
\\ \textsuperscript{1}Department of Mechanical and Industrial Engineering, \\ University of Iowa, Iowa City, IA 52242, USA 
\\\ \textsuperscript{2}Department of Radiology, Biomedical Engineering, \\ University of Iowa, Iowa City, IA 52242, USA 
\\\ \textsuperscript{3}Bioengineering Institute, \\ University of Auckland, Auckland, New Zealand \\}
\maketitle
\begin{abstract}
The airflow in a subject-specific breathing human lung is simulated with a multiscale computational fluid dynamics (CFD) lung model. The three-dimensional (3D) airway geometry beginning from the mouth to about 7 generations of airways is reconstructed from the multi-detector row computed tomography (MDCT) image at the total lung capacity (TLC). Along with the segmented lobe surfaces, we can build an anatomically-consistent one-dimensional (1D) airway tree spanning over more than 20 generations down to the terminal bronchioles, which is specific to the CT resolved airways and lobes (J Biomech 43(11): 2159-2163, 2010). We then register two lung images at TLC and the functional residual capacity (FRC) to specify subject-specific CFD flow boundary conditions and deform the airway surface mesh for a breathing lung simulation (J Comput Phys 244:168-192, 2013). The 1D airway tree bridges the 3D CT-resolved airways and the registration-derived regional ventilation in the lung parenchyma, thus a multiscale model. Large eddy simulation (LES) is applied to simulate airflow in a breathing lung (Phys Fluids 21:101901, 2009). In this fluid dynamics video, we present the distributions of velocity, pressure, vortical structure, and wall shear stress in a breathing lung model of a normal human subject with a tidal volume of 500 ml and a period of 4.8 s. On exhalation, air streams from child branches merge in the parent branch, inducing oscillatory jets and elongated vortical tubes. On inhalation, the glottal constriction induces turbulent laryngeal jet. The sites where high wall shear stress tends to occur on the airway surface are identified for future investigation of mechanotransduction.
\end{abstract}

\section*{}

The airflow in a subject-specific breathing human lung is simulated with a multiscale computational fluid dynamics (CFD) lung model. The three-dimensional (3D) airway geometry beginning from the mouth to about 7 generations of airways is reconstructed from the multi-detector row computed tomography (MDCT) image at the total lung capacity (TLC). Along with the segmented lobe surfaces, we can build an anatomically-consistent one-dimensional (1D) airway tree spanning over more than 20 generations down to the terminal bronchioles, which is specific to the CT resolved airways and lobes (Yin et al. 2010). We then register two lung images at TLC and the functional residual capacity (FRC) to specify subject-specific CFD flow boundary conditions and deform the airway surface mesh for a breathing lung simulation (Yin et al. 2013). The 1D airway tree bridges the 3D CT-resolved airways and the registration-derived regional ventilation in the lung parenchyma, thus a multiscale model. Large eddy simulation (LES) is applied to simulate airflow in a breathing lung (Choi et al. 2009). In this fluid dynamics video, we present the distributions of velocity, pressure, vortical structure, and wall shear stress in a breathing lung model of a normal human subject with a tidal volume of 500 ml and a period of 4.8 s. On exhalation, air streams from child branches merge in the parent branch, inducing oscillatory jets and elongated vortical tubes. On inhalation, the glottal constriction induces turbulent laryngeal jet. The sites where high wall shear stress tends to occur on the airway surface are identified for future investigation of mechanotransduction.
\\
\section*{References}

Choi, J., M.H. Tawhai, E.A. Hoffman, and C.-L. Lin, "On intra- and inter-subject variabilities of airflow in the human lungs," Phys. Fluids, 21:101901, 2009.

Yin, Y., J. Choi, E. A. Hoffman, M. H. Tawhai, and C.-L. Lin, "Simulation of pulmonary air flow with a subject-specific boundary condition," Journal of Biomechanics, 43(11):2159-2163, 2010.

Yin, Y., J. Choi, E. A. Hoffman, M. H. Tawhai, and C.-L. Lin, "A multiscale MDCT image-based breathing lung model with time-varying regional ventilation", Journal of Computational Physics, 244:168-192, 2013.

\end{document}